%% file: template.tex
\documentclass[cameraready]{Interspeech}
\usepackage{booktabs}
\usepackage{subcaption}
\usetikzlibrary{arrows.meta,positioning,calc}

\title{Neck-Learn: Attention-Based Multiple Instance Learning and Ensemble Framework for Ecological Momentary Assessment}

\author[affiliation={1,2}, orcid=0000-0002-5384-0973]{Ahsan Jamal}{Cheema}

\address{
    $^1$ Harvard University, Cambridge MA, USA\\
    $^2$ Eaton Peabody Laboratories, Massachusetts Eye and Ear Infirmary (MEEI), Boston, USA
}

\email{ahsancheema@g.harvard.edu}


\usepackage{comment}

\begin{document}
\maketitle
\begin{abstract}
Vocal hyperfunction (VH) is a prevalent voice disorder whose ambulatory detection remains challenging despite extensive daily voice data. Prior approaches capture week-long neck-surface accelerometer recordings but collapse them into fixed-length subject-level feature vectors, discarding within-day temporal dynamics encoding nuanced voicing feature interactions. We introduce a novel hybrid architecture combining gradient-boosted trees on day-level distributional features with a CNN-based multiple instance learning (MIL) framework that preserves and learns from from temporal dynamics throughout each day. On the held-out test set, our model exceeds the challenge baselines (AUC: 0.82 PVH, 0.77 NPVH), achieving AUCs of 0.879 for PVH (Rank 5) and 0.848 for NPVH (Rank 3), while also providing insights into clinically relevant information about both pathologies.

\end{abstract}

\input{1.Introduction/Introdution}

\input{2.Methods/Methods}

\input{3.Results/Results}

\input{4.Discussion/Discusssion}

\section{Generative AI Use Disclosure}
GenAI was used only to assist with editing and polishing the manuscript text. All scientific content, experimental design, code, data analysis, interpretation of results, intellectual contributions and manuscript drafts are solely the work of the authors. The authors reviewed and take full responsibility for the final content of this paper.

\bibliographystyle{IEEEtran}
\bibliography{references}

\end{document}

%% file: 1.Introduction/Introdution.tex
\section{Introduction}

Vocal hyperfunction (VH) is among the most common causes of voice disorders, affecting an estimated 7.6\% of the U.S.\ adult population at any given time \cite{Bhattacharyya2014-rp}. VH manifests in two clinically distinct forms: phonotraumatic VH (PVH), characterized by excessive laryngeal forces that lead to structural vocal fold lesions such as nodules and polyps, and nonphonotraumatic VH (NPVH), involving chronic vocal fatigue and discomfort without identifiable structural pathology \cite{VHframework2020, doi:10.1080/17549500802140153}. Accurate detection and differentiation of these subtypes is critical for guiding treatment. PVH often requires surgical intervention or intensive voice therapy, whereas NPVH management focuses on behavioral modification to improve vocal efficiency \cite{Van_Stan2021-oo, Gillespie2013-jn, Zhu2025-td}.

Traditional clinical assessment relies on in-clinic voice evaluation, which captures only a brief snapshot of vocal behavior under controlled conditions. This approach fails to account for the substantial day-to-day variability in voice use that occurs during daily activities \cite{Mehta2015-lw, Hunter2020}. Ambulatory voice monitoring using miniature neck-surface accelerometers addresses this limitation by enabling continuous, longitudinal recording of vocal behavior in ecological settings over days to weeks \cite{Cortes2018-plosone, Mehta2012-pv}. The accelerometer captures skin vibrations during phonation, from which acoustic and aerodynamic features can be derived via subglottal impedance-based inverse filtering (IBIF) \cite{Zanartu2013}, providing a rich characterization of voice use patterns while preserving the speech privacy.

Prior work on automated VH detection from ambulatory data has primarily employed subject-level summary statistics. Van Stan et al.\ \cite{Van_Stan2020-lk} demonstrated that higher-order distributional characteristics such as SPL skewness and H1--H2 variability reveal significant PVH group differences (AUC of 0.82--0.85), whereas average measures fail \cite{Mehta2015-lw, Popolo2005-st}. Ghassemi et al.\ \cite{Ghassemi2014} extended this with machine learning on distributional features for PVH classification. For NPVH, Van Stan et al.\ \cite{Van_Stan2021-oo} found that CPP mean and H1--H2 mode maximally differentiated NPVH from controls (AUC 0.78), reflecting a pathophysiological continuum of inefficient phonation. More recently, Cheema et al.\ \cite{Cheema2024-jvoice} demonstrated that relative fundamental frequency (RFF) a measure sensitive to vocal fold tension at voicing boundaries \cite{Heller_Murray2017-uv, Stepp2012-sv} can discriminate VH subtypes in ecological momentary assessments with accuracies of 81.3\% (PVH) and 62.5\% (NPVH). However, all prior approaches collapse temporal structure into fixed-length subject-level vectors to train simpler model due to absence of large scale data, or require specific unnnatural tasks, discarding potentially informative \textit{within-day} slow and fast dynamics such as progressive voice quality degradation, patterns of vocal loading, or transitions between high- and low-effort phonation segments.
The NeckVibe Challenge \cite{NeckVibe2026} provided a standardized benchmark for ambulatory VH detection, offering week-long accelerometer recordings from 582 subjects (213 PVH, 169 matched controls, 116 NPVH, 84 matched controls) with evaluation by subject-level area under the ROC curve (AUC). This represents the largest publicly available dataset for this task, with over 6,000 hours of neck-surface vibration data collected during daily life via a smartphone-connected accelerometer system \cite{Mehta2012-pv}.

In this work, we propose a hybrid ensemble that combines distributional feature engineering with deep temporal modeling via multiple instance learning (MIL); an approach not previously applied to ecological momentary assessment for voice biomarkers in general and vocal hyperfunction in particular. Our key contributions are:

\begin{enumerate}[leftmargin=*, itemsep=1pt, topsep=2pt]
    \item \textbf{Novel CNN-MIL architecture for ambulatory voice data:} We introduce a convolutional neural network with multi-head attention-based Multi Instance Learning (MIL) framework that processes variable-length sequences of window-level (seconds-scale) vocal features, treating each day-recording as a ``bag'' of instances. Unlike prior work that discards temporal structure \cite{Van_Stan2020-lk, Cortes2018-plosone, Ghassemi2014}, this approach learns dependencies and temporal relationships between windows throughout the day, identifying \textit{which} time segments are most discriminative for each VH subtype and learning the attention relationship between them throughpout the day.
    \item \textbf{Complementary dual-representation framework:} To account for inter-subject variability arising from different feature baselines and subject-level voice diversity, we pair higher-order distributional features, capturing global vocal behavior patterns informed by clinical evidence \cite{Van_Stan2020-lk, Van_Stan2021-oo}, with raw window-level time series that capture local temporal dynamics. These representations yield partially uncorrelated errors, enabling improved ensemble performance.
    \item \textbf{Strong empirical results on the largest VH dataset:} On the held-out NeckVibe test set, our ensemble achieves AUCs of 0.879 for PVH (Rank 5) and 0.848 for NPVH (Rank 3), substantially exceeding the challenge baselines and demonstrating particular strength on the clinically challenging NPVH task.
\end{enumerate}

%% file: 2.Methods/Methods.tex
\section{Methods}

\input{fig_combined}

\subsection{Data and Preprocessing}

The NeckVibe Challenge dataset \cite{NeckVibe2026} comprises week-long neck-surface accelerometer recordings from 582 subjects collected using a smartphone-based ambulatory voice monitor \cite{Cortes2018-plosone, Mehta2012-pv}. Each raw \texttt{.mat} file contains frame-level features at 50\,ms resolution for a single day-long recording (typically 10+ hours). We apply the \texttt{voiced\_speech} binary mask to retain only voiced speech frames, discarding silence and singing \cite{Van_Stan2020-lk, Mehta2015-lw}. From each masked frame, we extract 14 features organized into two groups:

\textit{Spectral/acoustic features (6):} Cepstral peak prominence (CPP), which measures voice periodicity and quality \cite{Van_Stan2021-oo, Patel2018}; accelerometer amplitude in dB~re~cm/s$^2$, reflecting vocal loudness; difference between the first and second harmonic magnitudes (H1--H2), reflecting glottal configuration and vocal fold closure patterns \cite{Van_Stan2020-lk}; low-to-high spectral power ratio; harmonic spectral tilt describing the slope of spectral energy decay; and estimated sound pressure level (SPL) at 15\,cm \cite{Popolo2005-st}.

\textit{IBIF-derived glottal airflow features (8):} Peak-to-peak glottal airflow amplitude (AC flow), closing quotient (CQ), IBIF H1--H2, harmonic richness factor (HRF), maximum flow declination rate (MFDR), normalized amplitude quotient (NAQ), open quotient (OQ), and speed quotient (SQ). These aerodynamic measures are derived from glottal airflow estimated via subglottal impedance-based inverse filtering (IBIF) of the neck-surface acceleration signal \cite{Zanartu2013}. Cortes et al.\ \cite{Cortes2018-plosone} demonstrated that these IBIF-derived features significantly differentiate PVH from controls (AUC 0.82, accuracy 0.83), motivating their inclusion.

\textit{Data cleaning:} IBIF inverse filtering commonly produces NaN and Inf artifacts due to numerical instability. We replace NaN$\rightarrow$0.0 and $\pm$Inf$\rightarrow$$\pm$1e5, ensuring stability while preserving valid frames.

\textit{Sliding window segmentation:} Voiced frames are divided into overlapping windows of \textbf{10 seconds} (200 frames at 50\,ms) with \textbf{5-second overlap}. Each window is compressed into a \textbf{56-dimensional} feature vector by computing 4 statistics per feature: mean, standard deviation, 5th, and 95th percentile ($14 \times 4 = 56$). Each day-recording is represented as a variable-length bag of 56-dim instance vectors of shape $(N_{\text{windows}}, 56)$.

\subsection{Feature Representation}

A key design choice is the use of two complementary feature representations (Fig.~\ref{fig:pipeline}), each designed to capture different aspects of vocal behavior.

\subsubsection{Subject-Level Distributional Features}

For each day-recording, we compute 11 distributional statistics across all windows for each of the 56 dimensional instance vectors as described earlier: mean, standard deviation, median, skewness, kurtosis, 5th/25th/75th/95th percentiles, IQR, and MAD. The inclusion of higher-order statistics is motivated by evidence that distributional characteristics not just central tendencies are critical for VH detection \cite{Van_Stan2020-lk, Ghassemi2014}. Two global features are appended: voiced frame ratio and total window count which implicitly captures how much voice does the individual typically use per day. This yields 618 features per day ($56 \times 11 + 2$). Day-level features are aggregated to the subject level by computing mean and standard deviation across days, plus the number of recording days, producing a \textbf{1237-dimensional} subject-level vector ($618 \times 2 + 1$). The cross-day standard deviation captures day-to-day variability, which may reflect inconsistent vocal patterns associated with hyperfunction \cite{Van_Stan2021-oo, Van_Stan2017-lp}.

\subsubsection{Window-Level Time Series (for CNN-MIL)}

The raw $(N_{\text{windows}}, 56)$ tensors are used directly, ensuring the sequential structure is preserved. This is the \textbf{key methodological novelty}: rather than collapsing temporal information into a fixed-length vector \cite{Van_Stan2020-lk, Cortes2018-plosone, Ghassemi2014}, we retain the full sequence and allow the model to learn temproal dynamics and relationships or attention weights between different windows throughout the day. Input features are normalized using a robust scaler (median and IQR from 30\% of training data) to reduce sensitivity to IBIF outliers.

\subsection{Cross-Validation and Data Splitting}

Since each subject contributes multiple day-recordings, naive random splitting risks placing different days from the same individual into both training and validation sets, a form of data leakage that Ghasemzadeh et al.\ \cite{Ghasemzadeh2024-ae} showed can overestimate accuracy by up to 50\% in speech and hearing ML studies. To prevent this, all models use \textbf{5-fold Stratified Group K-Fold} cross-validation with subject ID as the grouping variable. This ensures that \textit{all} recordings from a given participant remain in the same fold, so the model is never validated on data from a subject it saw during the training. Stratification preserves approximate class balance across folds. The primary evaluation metric is subject-level AUC computed on pooled out-of-fold (OOF) predictions across all 5 folds. 

\subsection{Model Architectures}
\textit{Note: Several other model architectures were tested and trained including RNN based models and contrastive learning based CNN models but the paper includes only the best performing model that was submitted as a part of the challenge.}
\subsubsection{Gradient-Boosted Trees}

Both XGBoost \cite{Breiman2001-tb} and LightGBM classifiers are trained on the 1237-dim subject-level vectors (Fig.~\ref{fig:pipeline}). They share: 500 estimators, max depth 5, learning rate 0.05, row/column subsampling at 80\%, L1/L2 regularization ($\alpha\ =\ 0.1$, $\lambda\ =\ 1.0$). To address class imbalance, the positive-class weight in the loss function is set to the ratio of negative to positive samples: 1.73 for PVH and 4.08 for NPVH. Early stopping (patience 50) prevents overfitting. Both are included for ensemble diversity through different tree-building algorithms (level-wise vs.\ leaf-wise growth).
\subsubsection{CNN with Multi-Head Attention MIL}
Each day-recording is treated as a ``bag'' of window ``instances'' under a MIL framework (Fig.~\ref{fig:cnnmil}):

\textit{CNN backbone:} Three 1D convolutional layers (kernel size 3, 128 filters) with GroupNorm (8 groups), ReLU, and dropout (0.4, 0.2). A residual connection on the third block that aids helps train faster. These CNN layers take the window level features and learn filter weights that can represent more abstract level features fromt he raw window level embeddings.
\textit{Multi-head attention pooling (4 heads):} Four parallel attention heads which take output from the CNN, independently compute softmax-weighted aggregations. Each has a linear layer ($128 \rightarrow\ 64$), Tanh, and scoring layer ($64 \rightarrow\ 1$). The four 128-dim outputs are concatenated into a 512-dim bag representation. Different heads can learn attention weights between different times during the same day and also learn different temporally abstract features that may be related to vocal hyperfunction behaviour.

\textit{Classifier head:} A 3-layer MLP ($512 \rightarrow\ 64 \rightarrow\ 32 \rightarrow\ 1$) with ReLU and dropout produces a single logit, trained with \texttt{BCEWithLogitsLoss} weighted by class imbalance.

\subsection{Ensemble Strategy}
Ensemble weights are optimized via exhaustive grid search (step 0.05) over all valid triplets ($w_{\text{CNN}} + w_{\text{XGB}} + w_{\text{LGB}} = 1$), maximizing AUC on pooled out-of-fold (OOF) predictions:
\begin{equation}
    p_{\text{final}} = w_1 \cdot p_{\text{CNN}} + w_2 \cdot p_{\text{XGB}} + w_3 \cdot p_{\text{LGB}}
\end{equation}
If the final $p_{\text{final}} \geq 0.5$ then the subject was marked as 1 (hyperfunction) otherwise 0 (normal). \\
As per challenge rules, when trainign/evaluating for PVH, both normal and NPVH were to be used as controls and when training/evaluating for NPVH both normal and PVH were to be used as controls, essentially giving use two different models (same architecture but different weights) for each type of pathology.

All models were trained and evaluated using \textbf{Software:} Python 3.11, PyTorch 2.5.1, scikit-learn 1.5.x, XGBoost 2.1.x, LightGBM 4.5.x, SciPy 1.14.x, NumPy 1.26.x, pandas 2.2.x. \textbf{Hardware:} Apple MacBook Pro with M-series chip (MPS backend). All hyperparameters are reported in Section~2.3 to enable full reproducibility.

%% file: fig_combined.tex
\begin{figure*}[t]
\centering
\vspace{-3mm}
\begin{minipage}[t]{0.48\textwidth}
\centering
\vspace{0pt}
\begin{tikzpicture}[
  every node/.style={font=\small, inner sep=3pt},
  block/.style={draw, rounded corners=2pt, fill=blue!5, minimum width=4.2cm, minimum height=0.55cm, align=center},
  model/.style={draw, rounded corners=2pt, fill=orange!10, minimum width=1.7cm, minimum height=0.5cm, align=center},
  ens/.style={draw, rounded corners=2pt, fill=green!10, minimum width=2.8cm, minimum height=0.55cm, align=center, font=\small\bfseries},
  out/.style={draw, rounded corners=2pt, fill=red!8, minimum width=2.4cm, minimum height=0.5cm, align=center},
  >=Stealth,
]

\node[block] (data) at (0,0) {Raw ACC Data (.mat)};
\node[block] (mask) at (0,-0.9) {Voiced Speech Masking};
\node[block] (win)  at (0,-1.8) {10s Windows (5s overlap)};
\node[block] (feat) at (0,-2.7) {56-dim Feature Vectors};

\draw[->,thick,black!60] (data) -- (mask);
\draw[->,thick,black!60] (mask) -- (win);
\draw[->,thick,black!60] (win) -- (feat);

\node[block, fill=blue!8, minimum width=2.6cm] (dist) at (-1.9,-4.3) {\begin{tabular}{c}Distributional\\Stats $\to$ 1237-d\end{tabular}};
\node[block, fill=blue!8, minimum width=2.6cm] (seq)  at (1.9,-4.3) {\begin{tabular}{c}Raw Sequences\\$(N\times\ 56)$\end{tabular}};

\draw[->,thick,black!60] (feat.south) -- ++(0,-0.5) -| (dist.north);
\draw[->,thick,black!60] (feat.south) -- ++(0,-0.5) -| (seq.north);

\node[font=\footnotesize\itshape, gray] at (-2.9,-3.4) {Global patterns};
\node[font=\footnotesize\itshape, gray] at (3.2,-3.4) {Temporal dynamics};

\node[model] (xgb) at (-3.0,-5.85) {XGBoost};
\node[model] (lgb) at (-0.4,-5.85) {LightGBM};

\draw[->,thick,black!60] (dist.south) -- ++(0,-0.35) -| (xgb.north);
\draw[->,thick,black!60] (dist.south) -- ++(0,-0.35) -| (lgb.north);

\node[model, minimum width=2.1cm] (cnn) at (2.0,-5.85) {\begin{tabular}{c}CNN-MIL\\(4-head Attn)\end{tabular}};
\draw[->,thick,black!60] (seq) -- (cnn);

\node[ens] (ensemble) at (0,-7.6) {Weighted Ensemble};

\draw[->,thick,black!60] (xgb.south) -- ++(0,-0.5) -| ([xshift=-10mm]ensemble.north);
\draw[->,thick,black!60] (lgb.south) -- ++(0,-0.35) -| ([xshift=-4.0mm]ensemble.north);
\draw[->,thick,black!60] (cnn.south) -- ++(0,-0.5) -| ([xshift=9mm]ensemble.north);

\node[out=45] (pred) at (0,-8.8) {VH Prediction (AUC)};
\draw[->,thick,black!60] (ensemble) -- (pred);

\node[font=\scriptsize, gray, left] at (-2.45,-6.8) {$w_1$};
\node[font=\scriptsize, gray, left] at (0.25,-6.6) {$w_2$};
\node[font=\scriptsize, gray, right] at (2.0,-6.8) {$w_3$};

\end{tikzpicture}
\subcaption{Pipeline architecture.}
\label{fig:pipeline}
\end{minipage}%
\hfill
\begin{minipage}[t]{0.48\textwidth}
\centering
\vspace{0pt}
\begin{tikzpicture}[
  every node/.style={font=\small, inner sep=3pt},
  block/.style={draw, rounded corners=2pt, fill=yellow!8, minimum width=4cm, minimum height=0.5cm, align=center},
  att/.style={draw, rounded corners=2pt, fill=purple!8, minimum width=0.95cm, minimum height=0.45cm, align=center, font=\footnotesize},
  mlp/.style={draw, rounded corners=2pt, fill=green!8, minimum width=1.8cm, minimum height=0.45cm, align=center},
  io/.style={draw, rounded corners=2pt, fill=blue!5, minimum width=2.4cm, minimum height=0.5cm, align=center},
  >=Stealth,
]

\node[io] (inp) at (0,0) {Bag: $N\times\ 56$};

\node[block] (c1) at (0,-0.85) {Conv1D $\to$ 128 + GN + ReLU};
\node[block] (c2) at (0,-1.7) {Conv1D $\to$ 128 + GN + ReLU};
\node[block] (c3) at (0,-2.55) {Conv1D $\to$ 128 + GN + ReLU};

\draw[->,thick,black!60] (inp) -- (c1);
\draw[->,thick,black!60] (c1) -- (c2);
\draw[->,thick,black!60] (c2) -- (c3);

\draw[->,thick,red!50,dashed,overlay] (c1.east) -- ++(0.8,0) |- node[right,pos=0.25,font=\footnotesize,red!60] {+residual} (c3.east);

\node[font=\footnotesize\itshape,gray,overlay] at (3.5,-3.0) {$N\times\ 128$ instance features};

\node[att] (h1) at (-1.5,-3.75) {Head 1};
\node[att] (h2) at (-0.5,-3.75) {Head 2};
\node[att] (h3) at (0.5,-3.75) {Head 3};
\node[att] (h4) at (1.5,-3.75) {Head 4};

\draw[->,thick,black!60] (c3.south) -- ++(0,-0.35) -| (h1.north);
\draw[->,thick,black!60] (c3.south) -- ++(0,-0.35) -| (h2.north);
\draw[->,thick,black!60] (c3.south) -- ++(0,-0.35) -| (h3.north);
\draw[->,thick,black!60] (c3.south) -- ++(0,-0.35) -| (h4.north);

\node[io] (cat) at (0,-4.85) {Concat $\to$ 512-dim};

\draw[->,thick,black!60] (h1.south) -- ++(0,-0.18) -| (cat.north west);
\draw[->,thick,black!60] (h2.south) -- ++(0,-0.12) -| ([xshift=-5.0mm]cat.north);
\draw[->,thick,black!60] (h3.south) -- ++(0,-0.12) -| ([xshift=5mm]cat.north);
\draw[->,thick,black!60] (h4.south) -- ++(0,-0.18) -| (cat.north east);

\node[mlp] (f1) at (0,-5.85) {FC 512 $\to$ 64};
\node[mlp] (f2) at (0,-6.65) {FC 64 $\to$ 32};
\node[mlp] (f3) at (0,-7.45) {FC 32 $\to$ 1};

\draw[->,thick,black!60] (cat) -- (f1);
\draw[->,thick,black!60] (f1) -- (f2);
\draw[->,thick,black!60] (f2) -- (f3);

\node[io, fill=red!8] (out) at (0,-8.3) {$\sigma$(probability)};
\draw[->,thick,black!60] (f3) -- (out);

\node[font=\footnotesize\itshape,gray,right,overlay] at (1.2,-6.65) {Classifier MLP};

\end{tikzpicture}
\subcaption{CNN-MIL architecture.}
\label{fig:cnnmil}
\end{minipage}
\caption{(a) Overall pipeline: raw accelerometer data is preprocessed into 56-dim window features, then processed via two paths (1) distributional statistics for tree models to learn global patterns and, (2) raw sequences for CNN-MIL to learn temporal dynamics and dependencies, which are combined through optimized ensemble weighting. (b) CNN-MIL: three Conv1D blocks with residual connection extract more abstract instance features; four attention heads compute softmax-weighted aggregations, concatenated and classified by a 3-layer MLP.}
\label{fig:architecture}
\end{figure*}
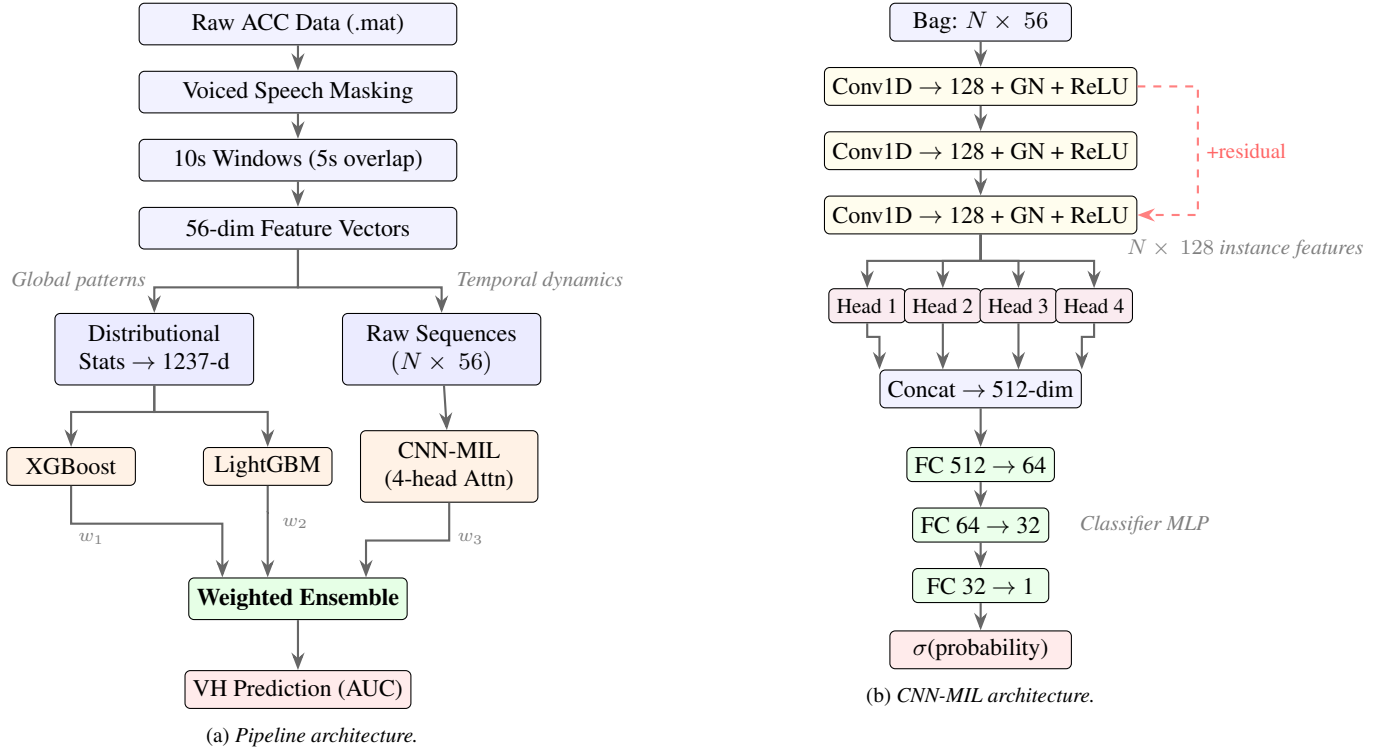

%% file: 3.Results/Results.tex
\section{Results}

\subsection{Individual and Ensemble Performance}

Table~\ref{tab:results} presents the mean out-of-fold (OOF) validation AUC for each model and the optimized ensemble on both classification tasks, alongside the held-out test set results from the NeckVibe Challenge leaderboard.

\begin{table}[t]
    \centering
    \caption{Mean 5-fold corss validated AUC on validation set. Bold indicates best results. Baseline values are from the challenge organizers \cite{NeckVibe2026}.}
    \label{tab:results}
    \begin{tabular}{lcccc}
        \toprule
        & \multicolumn{2}{c}{\textbf{OOF (Val)}} & \multicolumn{2}{c}{\textbf{Test Set}} \\
        \cmidrule(lr){2-3} \cmidrule(lr){4-5}
        \textbf{Model} & \textbf{PVH} & \textbf{NPVH} & \textbf{PVH} & \textbf{NPVH} \\
        \midrule
        XGBoost             & 0.845 & 0.601 & -- & -- \\
        LightGBM            & 0.824 & 0.671 & -- & -- \\
        CNN--Attn MIL       & 0.845 & 0.765 & -- & -- \\
        \midrule
        \textbf{Ensemble}   & \textbf{0.880} & \textbf{0.770} & \textbf{0.879} & \textbf{0.848} \\
        \textit{Baseline}   & \textit{--} & \textit{--} & 0.82 & 0.78 \\
        \bottomrule
    \end{tabular}
\end{table}

The optimized ensemble weights for PVH are: CNN-MIL 0.45, XGBoost 0.35, LightGBM 0.20. For NPVH: CNN-MIL 0.50, XGBoost 0.15, LightGBM 0.35. The higher CNN-MIL weight for NPVH reflects its substantially better individual performance on this more challenging task, where temporal patterns captured by the MIL framework provide information that distributional summaries miss. On the held-out test set, our ensemble achieves PVH AUC of 0.879 (Rank 5) and NPVH AUC of 0.848 (Rank 3), demonstrating strong generalization. Notably, the NPVH test performance (0.848) substantially exceeds the challenge baseline (0.78), suggesting that the ensemble's temporal modeling captures generalizable patterns of inefficient phonation that transfer well to unseen subjects.

\subsection{Utility of Incorporating Temporal Framework}

To rigorously evaluate the contribution of each component, Table~\ref{tab:ablation} presents a systematic ablation where models are evaluated individually and in pairwise combinations.

\begin{table}[t]
    \centering
    \caption{Ablation on Ensembles: Mean OOF AUC for individual models and pairwise combinations. $\Delta$ shows improvement of the full ensemble over the best single model.}
    \label{tab:ablation}
    \begin{tabular}{lcc}
        \toprule
        \textbf{Configuration} & \textbf{PVH} & \textbf{NPVH} \\
        \midrule
        XGBoost only                & 0.845 & 0.601 \\
        LightGBM only               & 0.824 & 0.671 \\
        CNN-MIL only                & 0.845 & 0.765 \\
        \midrule
        XGB + LGB (equal)           & 0.853 & 0.658 \\
        XGB + CNN-MIL (equal)       & 0.867 & 0.721 \\
        LGB + CNN-MIL (equal)       & 0.856 & 0.740 \\
        \midrule
        \textbf{Full Ensemble (opt.)} & \textbf{0.880} & \textbf{0.770} \\
        \midrule
        $\Delta$ vs.\ best single   & +0.035 & +0.005 \\
        \bottomrule
    \end{tabular}
\end{table}

Three findings stand out. First, the CNN-MIL model provides the largest individual contribution, particularly for NPVH (0.765 vs.\ 0.601--0.671 for tree models), confirming that temporal dynamics especially on smaller timescales of 10 sec are crucial for NPVH detection. Second, combining any two model types consistently outperforms either alone for PVH, demonstrating that distributional and temporal features capture partially orthogonal information. Third, the optimized three-model ensemble outperforms all pairwise combinations, validating the value of ensemble diversity from different tree-building algorithms (level-wise XGBoost vs.\ leaf-wise LightGBM) and combining with MIL framework.

%% file: 4.Discussion/Discusssion.tex
\section{Discussion}

Voice is increasingly recognized as a biomarker where temporal dynamics not just static acoustic snapshots carry diagnostic information across conditions from Parkinson's disease to depression \cite{Verde2021-voicebiomarker}. Yet ambulatory VH detection has previously relied exclusively on collapsing days of vocal data into fixed-length summary vectors. Our results demonstrate that this could discard clinically meaningful signal, and that a MIL framework reasoning over within-day trajectories yields substantial gains, particularly for NPVH.

The theoretical basis lies in the physiology of vocal loading. Titze et al.\ \cite{Titze2003-np} formalized \textit{vocal dose}, the accumulated vibratory exposure of vocal fold tissue, showing that tissue damage follows a cumulative dose-response relationship rather than instantaneous behavior. Hunter and Titze \cite{Hunter2009-recovery} further demonstrated that vocal fatigue recovery follows characteristic temporal trajectories with recovery over 4--6 hours, implying that the \textit{pattern} of loading and rest throughout a day is as diagnostically informative as overall voice use level. For NPVH patients, who experience chronic vocal fatigue without structural lesions \cite{doi:10.1080/17549500802140153}, these temporal loading patterns are likely the primary signature distinguishable from healthy speakers explaining why our CNN-MIL attention based architecture achieves substantially higher NPVH performance than distributional models that average away these trajectories.

The MIL formulation suits this problem because diagnostic labels exist at the subject level, but discriminative evidence is distributed unevenly across the day. The attention mechanism \cite{Ilse2018-attention} learns which instances matter most and has proven effective in computational pathology \cite{Campanella2019-pathology} and ambient speech analysis from wearables \cite{Xu2020-speech-mil}. The 4-head architecture may enable specialization: different heads could weight periods of voice feature variance and changes throughout the day, consistent with hypothesis by \cite{Titze2003-np}. This differential attention explains the clinically coherent PVH--NPVH contrast: PVH's structural lesions produce acoustic signatures that get repeated more consistently and are captured well by that distributional (or average) statistics, while NPVH's functional inefficiency manifests through temporal patterns i.e. progressive voice quality degradation, inconsistent phonatory effort, and atypical fatigue-recovery dynamics \cite{Hunter2020, Van_Stan2021-oo} that only temporal modeling can exploit.

\subsection{Limitations and Future Work}

Our study has several limitations. The CNN-MIL processes each day independently; a hierarchical model jointly reasoning across days could capture week-level temporal trends in vocal behavior. While the gradient boosting methods are more interpretable, the CNN based methods have large number of features, which makes them difficult to interpret however, attention visualization, which could yield some useful clinical insights on a more abstract level nonlinearly combined features that drive the temporal dynamics and VH detection. In the current study we only use the voicing segments throughout the day, future work can exploit the transition between voiced and unvoiced segments to improve performance of the models. Indeed RFF, a measure derived from transition between voiced and unvoiced phonemes, has shown promise for discriminating VH subtypes \cite{Cheema2024-jvoice, Heller_Murray2017-uv, Stepp2012-sv, app12168121}. However, given the fact that our model captures temporal dynamics so it has capability to model vocal strain or vocal effort as well which has proven elusive to model even with RFF based measures \cite{Cheema2024-jvoice}. Another limitation is that the model is not causal i.e. it learns to make prediction after seeing the entire day/multiple day dataset from an individual, future work could explore developing models where only data from past timestamps are used to make predictions for a current observation \cite{10253331, Wang2025-lz, akbar2025an, Wang2023-sx}.